\documentstyle[epsfig]{cjp3}
\def\nn{\nonumber}
\begin{document}
\hyphenation{brems-strah-lung}
\hyphenation{pa-ra-me-tri-zed}
\title{Virtual Compton Scattering off the Pseudoscalar Meson Octet}
\authori{T. Fuchs,$^{\rm a}$\, B. Pasquini,$^{{\rm b}}$ \, 
C. Unkmeir,$^{{\rm a}}$ and S. Scherer$^{{\rm a}}$}

\addressi{$^{{\rm a}}$Institut f\"ur Kernphysik, Johannes Gutenberg-Universit\"at,
D-55099 Mainz, Germany\\
$^{{\rm b}}$ ECT$^*$, European Centre for Theoretical Studies in
Nuclear Physics and Related Areas, I-38050 Villazzano (Trento), Italy
}

\authorii{}       
\addressii{}
\authoriii{}      
\addressiii{}     

\headtitle{Virtual Compton Scattering off the Pseudoscalar Meson Octet}
\headauthor{T. Fuchs, B. Pasquini, C. Unkmeir, and S. Scherer}

\evidence{A}
\daterec{XXX}    
\cislo{0}  \year{2000}
\setcounter{page}{1}

\maketitle

\begin{abstract}
We present a calculation of the virtual Compton scattering amplitude 
for the pseudoscalar meson octet in the framework of chiral
perturbation theory at ${\cal O}(p^4)$. 
We calculate the electromagnetic generalized polarizabilities 
and compare the results in the real Compton scattering limit 
to available experimental values.
Finally, we give predictions  for the differential cross section 
of electron-meson bremsstrahlung.
\end{abstract}

\section{Introduction}
Virtual Compton scattering (VCS) is a suitable tool to obtain
information on the internal structure of a system.
Compared to real Compton scattering (RCS), the VCS reaction contains
much more information because of the independent variation of energy
and momentum transfer to the target and the additional longitudinal 
polarization of the virtual photon.
In particular, VCS off the pseudoscalar meson octet ($\pi$, K, $\eta$) is of
theoretical interest, since the pseudoscalar mesons can be identified with the
Goldstone bosons of QCD which arise from the spontaneous symmetry
breakdown from ${\rm SU}(3)_{\rm L} \times {\rm SU}(3)_{\rm R}$ to
${\rm SU}(3)_{\rm V}$.

For a low-energy outgoing real photon, the VCS amplitude 
can be parametrized in terms of generalized polarizabilities \cite{GLT}
which give information about the response of the target to a
quasi-static electromagnetic field.
These observables are functions of the square of the virtual photon
four-momentum, $q^2$, and reduce in the limit of $q^2\rightarrow
0$  to the standard electric ($\alpha$) and magnetic ($\beta$) 
polarizabilities  of real Compton scattering.

The VCS amplitude of mesons can be investigated experimentally
through inelastic scattering of mesons ($M$) off atomic electrons, 
$ M + e  \to e + M + \gamma$.
In the case of the pion, such an experiment has been performed by the
SELEX collaboration at Fermilab \cite{Selex}. Other experiments to
determine the pion polarizabilities are presently under analysis or 
planned at several facilities \cite{exp1,exp2,exp3}. 
In addition, experiments to measure the kaon polarizabilities have
been proposed at CERN \cite{Compass}.

In this paper, we investigate the VCS reaction for pions and kaons 
in the framework of chiral perturbation theory (ChPT) at ${\cal O}(p^4)$.
In Sec.~\ref{section2}, we give a short survey of ChPT,  
and in Sec.~\ref{section3} we discuss our predictions for the VCS amplitude.
In Sec.~\ref{section4}, we present our results for the generalized
pion and kaon polarizabilities in comparison with different model calculations.
In Sec.~\ref{section5} we show predictions
for the cross section of electron-pion bremsstrahlung, 
discussing the possibility to extract information about the
generalized polarizabilities.
Finally, a short summary and some conclusions are given in Sec.~\ref{section6}.

\section{Chiral Perturbation Theory}
\label{section2}
We assume that the ${\rm SU}(3)_{\rm L}\times {\rm SU}(3)_{\rm R}$ symmetry
of the QCD Lagrangian in the chiral limit, i.e., vanishing u-, d-, and
s-quark masses, is spontaneously broken down 
to ${\rm SU}(3)_{\rm V}$ \cite{GL1, GL2}, giving rise to
eight massless pseudoscalar Goldstone bosons with vanishing interactions in
the limit of zero energy \cite{Goldstone}.
These Goldstone bosons can be identified with the pseudoscalar
meson octet, the non-zero mass resulting from an explicit symmetry
breaking in QCD through the quark masses.
   
The interactions of the pseudoscalar mesons can be described using the
effective chiral Lagrangian 
${\cal L} = {\cal L}_2 + {\cal L}_4 + \ldots,$
where the subscript $2n$ refers to the order in the so-called momentum
expansion.
In such an expansion we count derivatives as ${\cal O}(p)$ and quark masses
as ${\cal O}(p^2)$. The coupling to the external electromagnetic field 
and the explicit symmetry breaking due to the quark
masses are included as perturbations.
Then the lowest order Lagrangian can be written as
  \begin{equation}
    \label{eq:l2}
      {\cal L}_2 = \frac{F_0^2}{4}
                  \mbox{Tr}\left[ D_{\mu}U \left(D^{\mu}U\right)^{\dagger} +
                  \chi U^{\dagger} + U \chi^{\dagger} \right],
  \end{equation}
and the next to leading order Lagrangian reads
(see Eq. (6.16) of Ref. \cite{GL2})
  \begin{eqnarray} 
    \label{eq:l4}
    {\cal L}_4 &=& \ldots -
       {\rm i} L_9\mbox{Tr}\left(
         F^{R}_{\mu\nu}D^{\mu}U\left(D^{\nu}U\right)^{\dagger} +
         F^{L}_{\mu\nu}\left(D^{\mu}U\right)^{\dagger}D^{\nu}U\right)\nn\\ &&
       + L_{10}\mbox{Tr}
         \left(UF^{L}_{\mu\nu}U^{\dagger}F_{R}^{\mu\nu}\right) + \ldots.
  \end{eqnarray}
In Eqs. (\ref{eq:l2}) and (\ref{eq:l4}), $U$ is given by
  \begin{displaymath}
     U(x) = \exp{\left(\frac{{\rm i}\Phi(x)}{F_0}\right)},
  \end{displaymath}
where $F_0$ is the pion decay constant in the
chiral limit, $F_\pi = F_0(1 + {\cal O}(p^2)) = 92.4\, \mbox{MeV}$ and
$\Phi(x)$ is given by
  \begin{displaymath}
     \Phi(x) = \left(\begin{array}{ccc}
           \pi^0 + \frac{1}{\sqrt{3}}\eta & \sqrt{2}\pi^{+} & \sqrt{2}K^{+} \\
           \sqrt{2}\pi^{-} & -\pi^0 + \frac{1}{\sqrt{3}}\eta & \sqrt{2}K^0 \\
           \sqrt{2}K^{-} & \sqrt{2}\bar{K}^0 & -\frac{2}{\sqrt{3}}\eta
               \end{array}\right).
  \end{displaymath}
 
The quantity $\chi$ contains the quark masses,
$\chi=2B_0\,\mbox{diag}(m,m,m_{s})$, where we assumed perfect isospin symmetry,
$m_u = m_d = m$. The constant $B_0$ is related to the quark condensate
$\langle\bar{q}q\rangle$. The coupling to the electromagnetic field is
contained in the covariant derivative $D^\mu$ 
and the field strength tensor $F^{\mu\nu}$,
defined respectively by
  \begin{eqnarray*}
     D_{\mu}U &=& \partial_{\mu}U+{\rm i}eA_{\mu}\left[Q,U\right],\\
     F^{R}_{\mu\nu} = F^{L}_{\mu\nu} &=&
        -eQ\left(\partial_{\mu}A_{\nu}-\partial_{\nu}A_{\mu}\right),
  \end{eqnarray*}
with $Q = \mbox{diag}\left(2/3,-1/3,-1/3\right).$
Finally, using Weinberg's power counting scheme \cite{Weinberg} 
one can classify the contribution of Feynman diagrams in the momentum 
expansion. 

\section{VCS amplitude}
\label{section3}

\begin{figure}[ht]
   \begin{center}
     \mbox{\epsfig{file=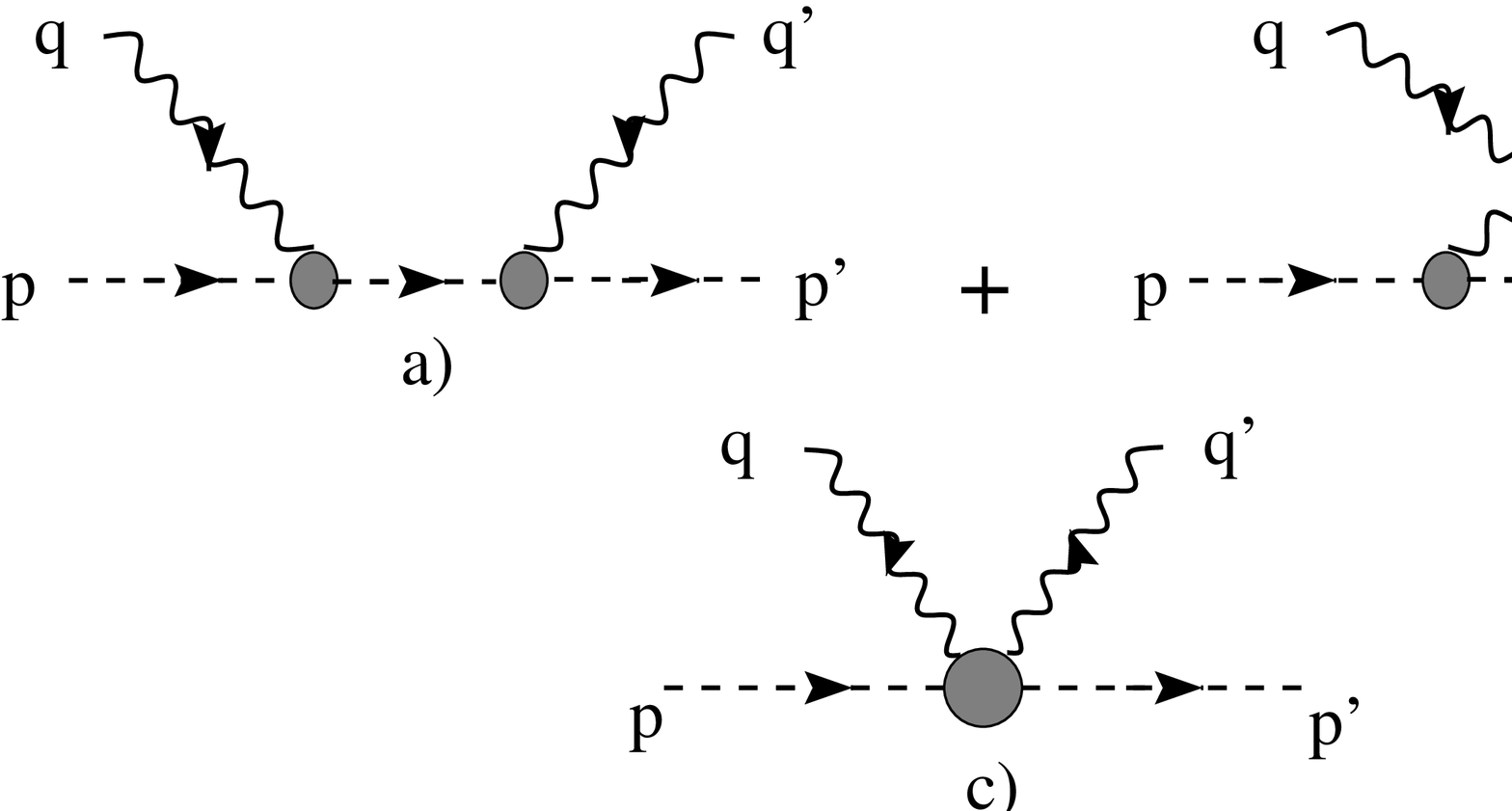, height=3.0cm, width=10.cm}}      
   \end{center}
\caption{\label{fig:born_nb} Compton scattering amplitude:
${\rm a})$ and ${\rm b})$ pole graphs;
${\rm c})$ one-particle-irreducible contribution. 
All the building blocks are renormalized.}
\end{figure}

In this section we discuss the VCS amplitude for pions,
$\gamma^\ast(q, \epsilon) + \pi(p) \to
\gamma(q^\prime, \epsilon^\prime) + \pi(p^\prime)$, and kaons,
$\gamma^\ast(q, \epsilon) + K(p) \to
\gamma(q^\prime, \epsilon^\prime) + K (p^\prime)$,
 with an outgoing real photon ($q^{\prime\,2}=0$). 
Throughout the calculation we use the conventions of Bjorken and 
Drell \cite{BD}, except $e>0$. 
Since the amplitude for the neutral particles is contained in the
amplitude for the charged particles, we can write
  \begin{eqnarray*}
    {\cal M}^{\rm VCS}_{\pi^{\pm}} &=& {\cal A}+{\cal B},\qquad
    {\cal M}^{\rm VCS}_{\pi^0} = {\cal A}, \\
    {\cal M}^{\rm VCS}_{K^{\pm}} &=& {\cal C}+{\cal D}, \qquad
    {\cal M}^{\rm VCS}_{K^0} = {\cal M}^{\rm VCS}_{\bar{K}^0}={\cal C}.
  \end{eqnarray*}
In addition, we split ${\cal M}^{\rm VCS}$ into a Born (B) and a non-Born (NB)
term which we construct such that they are separately gauge invariant,
${\cal M}^{\rm VCS} = {\cal M}_{\rm B}^{\rm VCS} +
{\cal M}_{\rm NB}^{\rm VCS}$
(see Fig.~\ref{fig:born_nb}).

\noindent The result for the gauge-invariant generalized Born terms reads
  \begin{eqnarray}
     {\cal B}_{\rm B} &=& -2{\rm i}e^2 F_{\pi}(q^2)\left[\frac{
     \epsilon\cdot (2p+q) \epsilon'\,^\ast \cdot p'}{s-m_\pi^2}
     + \frac{\epsilon\cdot (2p'-q) \epsilon'\,^\ast \cdot p}{u-m_\pi^2}
     - \epsilon\cdot\epsilon'\,^\ast\right],  \\
     {\cal D}_{\rm B} &=& -2{\rm i}e^2 F_{K}(q^2)\left[\frac{
     \epsilon\cdot (2p+q) \epsilon'\,^\ast \cdot p'}{s-m_K^2}
     + \frac{\epsilon\cdot (2p'-q) \epsilon'\,^\ast \cdot p}{u-m_K^2}
     - \epsilon\cdot\epsilon'\,^\ast\right],
  \end{eqnarray}
where $F_{\pi(K)}$ is the charged pion (kaon) form factor.
\noindent  The gauge-invariant residual amplitude is given by
  \begin{eqnarray}
     \label{eq:residual_ampl1}
     {\cal A}_{\rm NB} &=& -{\rm i}e^2
     (q'\cdot\epsilon q\cdot \epsilon'\,^\ast-q\cdot q'
     \epsilon\cdot\epsilon'\,^\ast)
     \frac{1}{8\pi^2F_0^2 q\cdot q'}
     \left[(m^2_\pi-t){\cal G}_{\pi} - \frac{1}{4}t
     {\cal G}_{K}\right], \\
     {\cal B}_{\rm NB} &=& -{\rm i}e^2
     (q'\cdot\epsilon q\cdot \epsilon'\,^\ast-q\cdot q'
     \epsilon\cdot\epsilon'\,^\ast)
     \left[-\frac{8(L_9^r+L_{10}^r)}{F_0^2}
     - \frac{2 m^2_\pi-t}{16 \pi^2 F_0^2q\cdot q'}{\cal G}_{\pi}
     \right], \\
     {\cal C}_{\rm NB} &=& {\rm i}e^2
     (q'\cdot\epsilon q\cdot \epsilon'\,^\ast-q\cdot q'
     \epsilon\cdot\epsilon'\,^\ast)
     \frac{1}{16\pi^2F_0^2 q\cdot q'}
     \frac{t}{2}\left({\cal G}_{\pi} + {\cal G}_{K}
     \right), \\
     \label{eq:residual_ampl2}
     {\cal D}_{\rm NB} &=& -{\rm i}e^2
     (q'\cdot\epsilon q\cdot \epsilon'\,^\ast-q\cdot q'
     \epsilon\cdot\epsilon'\,^\ast)
     \left[-\frac{8(L_9^r+L_{10}^r)}{F_0^2}
     - \frac{t}{32\pi^2 F_0^2 q\cdot q'}
     {\cal G}_{K}\right],
\end{eqnarray}
where $s$, $t$, and $u$ are the standard Mandelstam variables. 
In Eqs. (\ref{eq:residual_ampl1}) - (\ref{eq:residual_ampl2})
the function 
${\cal G}_{\pi (K)} := {\cal G}(m^2_{\pi (K)},q^2,q \cdot q')$
corresponds to the pion (kaon) loop integral contributions, and is
explicitly given in the appendix.
We note that out of the 12 low-energy constants contained 
in ${\cal L}_4$ only the combination 
$L^r_9 + L^r_{10}$ enters in our result, with
the value $L^r_9 + L^r_{10} = (1.43\pm 0.27)\times 10^{-3}$ being 
determined through the decay $\pi^+\to e^+\nu_e\gamma$.

\section{Generalized Polarizabilities}
\label{section4}
Following the analysis of Ref. \cite{USLD}, 
the invariant VCS amplitude can be parametrized as
  \begin{equation}
    \label{invam}
    - {\rm i}{\cal M} = B_1F^{\mu\nu}F'_{\mu\nu} + \frac{1}{4} B_2
               \left(P_\mu F^{\mu\nu}\right) \left(P^\rho F'_{\rho\nu}\right)
               + \frac{1}{4}B_3 \left(P^\nu q^\mu F_{\mu\nu} \right)
                 \left(P^\sigma q^\rho F'_{\rho\sigma} \right),
  \end{equation}
where $P_\mu = p_\mu + p'_\mu$,
$F^{\mu\nu} = -{\rm i}(q^\mu \varepsilon^\nu - q^\nu \varepsilon^\mu)$ and
$F'_{\mu\nu} = {\rm i}(q'_\mu \varepsilon'\,^*_\nu -
q'_\nu \varepsilon'\,^*_\mu)$.
The structures of Eq. (\ref{invam}) are particularly simple when evaluated
in the pion Breit frame (PBF) defined by $\bm P = \bm 0$. In this frame, the
structures $\bm{E}_L\cdot \bm{E}'$, $\bm{E}_T\cdot \bm{E}'$, and
$\bm{B}\cdot \bm{B}'$ appear.
We now consider the limit $\omega^\prime \to 0$ of the non-Born amplitude
${\cal M}_{\rm NB}$, for which $B_{i\,{\rm NB}} \to b_{i\,{\rm NB}}$, and
define three generalized dipole polarizabilities, 
  \begin{eqnarray}
     8\pi m_\pi \beta(q^2) &\equiv& b_{1\,{\rm NB}}, \\
     8\pi m_\pi \alpha_T(q^2) &\equiv& -b_{1\,{\rm NB}}(q^2) -
                \left(m_\pi^2 - \frac{q^2}{4}\right) b_{2\,{\rm NB}}(q^2), \\
     8\pi m_\pi\alpha_L(q^2) &\equiv& -b_{1\,{\rm NB}}(q^2) -
                \left(m_\pi^2 - \frac{q^2}{4}\right) \left[
                b_{2\,{\rm NB}}(q^2) + q^2 b_{3\,{\rm NB}}(q^2)\right].
  \end{eqnarray}
We note that $[\bm B\cdot \bm B^\prime]_{\rm PBF}$ and
$[\bm{E}_L\cdot \bm{E}^\prime]_{\rm PBF}$ are of ${\cal O}(\omega^\prime)$
whereas $[\bm{E}_T\cdot \bm{E}^\prime]_{\rm PBF}={\cal O}(\omega'^2)$, i.e.,
that different powers of $\omega^\prime$ have been kept.

In ChPT up to ${\cal O}(p^4)$, $b_{2\,{\rm NB}}(q^2) = b_{3\,{\rm NB}}(q^2)=0$,
which leads to the relation
  \begin{equation}
    \alpha_L(q^2) = \alpha_T(q^2) = -\beta(q^2).
  \end{equation}
The explicit expressions for the polarizabilities read
  \begin{eqnarray}
    \label{polpm}
    \alpha_L^{\pi^\pm}(q^2) &=&
    \frac{e^2}{8\pi m_\pi}
    \left\{\frac{8\left(L_9^r+L_{10}^r\right)}{F_0^2}\right. \nn \\ && \left. 
    - \frac{q^2}{m_\pi^2}\frac{1}{\left(4\pi F_0\right)^2}
      \left[J^{(0)'}\left(\frac{q^2}{m^2_\pi}\right)
      + \frac{1}{2}\frac{m^2_\pi}{m^2_K}
      J^{(0)'}\left(\frac{q^2}{m^2_K}\right)\right]
    \right\},\\
    \label{polp0}
    \alpha_L^{\pi^0}(q^2) &=&
    \frac{e^2}{4\pi} \frac{1}{\left(4\pi F_0\right)^2 m_\pi}
      \left[-\frac{1}{4}\frac{q^2}{m_k^2}
      J^{(0)'}\left(\frac{q^2}{m^2_K}\right)
      + \left(1-\frac{q^2}{m_\pi^2}\right)
      J^{(0)'}\left(\frac{q^2}{m_\pi^2}\right)\right], \nn \\ &&\\
    \label{polkp}
    \alpha_L^{K^\pm}(q^2) &=&
    \frac{e^2}{8\pi m_K}
    \left\{\frac{8\left(L_9^r+L_{10}^r\right)}{F_0^2} \right. \nn \\ && \left.
    - \frac{q^2}{m_K^2}\frac{1}{\left(4\pi F_0\right)^2}
      \left[\frac{1}{2}\frac{m^2_K}{m^2_\pi}
      J^{(0)'}\left(\frac{q^2}{m^2_\pi}\right)
      + J^{(0)'}\left(\frac{q^2}{m^2_K}\right)\right]
    \right\},\\
    \label{polk0}
    \alpha_L^{K^0}(q^2) &=&
    -\frac{e^2}{4\pi} \frac{1}{\left(4\pi F_0\right)^2 }
    \frac{q^2}{4m_K^3}
      \left[\frac{m_K^2}{m_\pi^2}
      J^{(0)'}\left(\frac{q^2}{m_\pi^2}\right)
      + J^{(0)'}\left(\frac{q^2}{m^2_K}\right)\right],
  \end{eqnarray}
where the function $J^{(0)'}(x)$ is given in the appendix. Note that the
${\cal L}_4$ low-energy constants enter only in the result for the charged
particles. 

\begin{figure}[ht]
  \begin{center}
     \mbox{\psfig{file=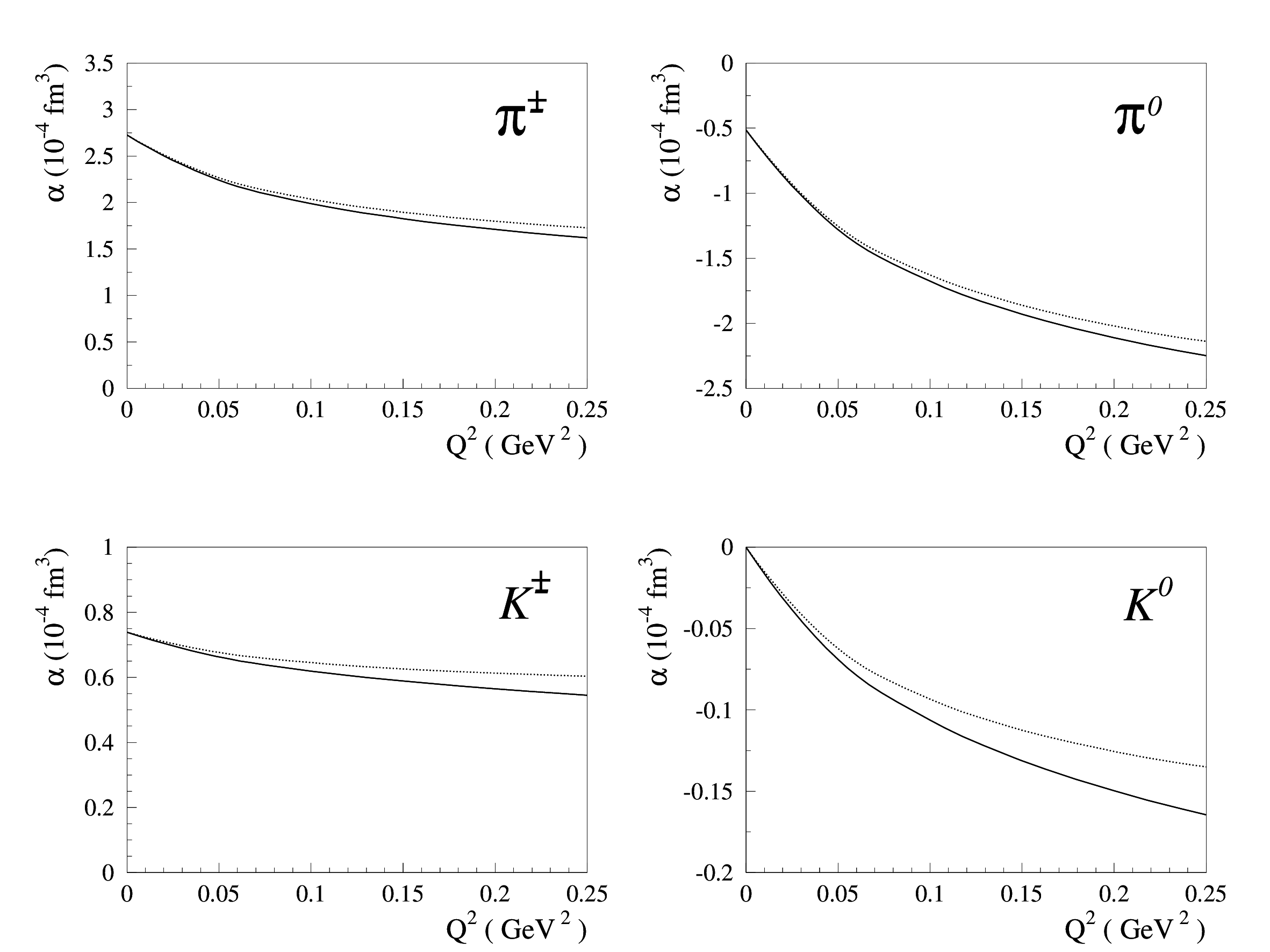,height=7.5cm, width=11cm}} 
  \end{center}
\caption{\label{pol} ${\cal O}(p^4)$ predictions for the generalized
pion and kaon polarizabilities $\alpha_L(q^2=-Q^2)$.
The solid lines correspond to the total results, while 
the dotted lines result from pion-loop contributions, only.}
\end{figure}

The results for the generalized polarizabilities as function of $Q^2=-q^2$
are shown in Fig. \ref{pol}. We made use of the numerical value
$F_0 \simeq F_\pi = 92.4 \, \mbox{MeV}$.

At $q^2=0$, Eqs. (\ref{polpm})-(\ref{polk0}) reduce to the RCS
polarizabilities given by
  \begin{eqnarray}
     \alpha_{\pi^{\pm}} = -\beta_{\pi^{\pm}} &=&
     (2.72\pm0.42)\times 10^{-4}\,\mbox{fm}^3,\\
     \alpha_{\pi^0} = -\beta_{\pi^0} &=& -0.50 \times 10^{-4}\,\mbox{fm}^3,\\
     \alpha_{K^{\pm}} = -\beta_{K^{\pm}} &=&
     (0.74\pm0.12)\times 10^{-4}\,\mbox{fm}^3,\\
    \alpha_{K^0\,(\bar{K}^0)} = \beta_{K^0\, (\bar{K}^0)} &=& 0.
  \end{eqnarray}

In Tabs. \ref{exp} and 2 we collect the available experimental
and theoretical results for the RCS polarizabilities. 
We note that the ChPT predictions are remarkably smaller compared to
the experimental values and the other theoretical results.

\begin{table}[t]
  \begin{center}
  \begin{tabular}{|c||c|c|}
    \hline
    reaction & $\gamma p \rightarrow \gamma \pi^+ n$ &
    $\pi^-A\rightarrow \pi^-A\gamma$ \\ 
    \hline 
    $\alpha_{\pi^{\pm}}$ & $20\pm 12\,$ \cite{Aib} 
    & $6.8\pm 1.4\,$ \cite{Ant1}
    \\ \hline
    $\beta_{\pi^{\pm}}$ & $-$ &
    $-7.1 \pm 2.8 \pm 1.8\,$ \cite{Ant2} \\ \hline
    $\alpha_{\pi^{\pm}}+\beta_{\pi^{\pm}}$ & $-$ &
    $1.4 \pm 3.1 \pm 2.5\,$ \cite{Ant2}
    \\ \hline
  \end{tabular}
  \end{center}
\caption{\label{exp} Experimental predictions for the pion polarizabilities in
units of $10^{-4}\,\mbox{fm}^3$.}
\end{table}

\begin{table}[t]
  \begin{center}
  \begin{tabular}{|c||c|c|c|c|c|c|}
     \hline
     & chiral QM & nonrel. QM & NJL & ChPT & QCM\\
     \hline
     $\alpha_{\pi^{\pm}}$
     & $8.0$ & $0.05$ & $10.5 - 12.5$ & $2.4\pm 0.5$ &  $3.63$\\ \hline
     $\beta_{\pi^{\pm}}$  
     & $-7.8$ & $-$ & $-(10.3 - 11.8)$ & $-2.1\pm 0.5$ & $-2.41$\\ \hline
     $\alpha_{\pi^0}$ 
     & $-$ & $-$ & $-$ & $-0.35\pm 0.10$ & $0.74$\\ \hline
     $\beta_{\pi^0}$
     & $-$ & $-$ & $-$ & $1.50\pm 0.20$ & $-0.30$\\ \hline
     $\alpha_{K^{\pm}}$  & $9.7$ & $-$ & $-$ &
     $0.64 \pm 0.1$ & $2.28$\\ \hline
     $\beta_{K^{\pm}}$ & $-4.4$ & $-$ & $-$ & $-$ & $-1.31$ \\ \hline
  \end{tabular}
  \end{center}
\caption{\label{theo} Theoretical predictions for the pion and kaon 
polarizabilities (in units of $10^{-4}\,\mbox{fm}^3$) 
within different models, like
chiral quark model (QM) \cite{chiral},
nonrelativistic quark model (nonrel. QM)  \cite{Sch}, 
Nambu-Jona-Lasinio (NJL) model \cite{Ber}, 
ChPT \cite{chpt}, and 
quark confinement model (QCM) \cite{Iva}.}
\end{table}

\section{Electron-Meson Bremsstrahlung}
\label{section5}
\begin{figure}[ht]
   \begin{center}
      \mbox{\psfig{file=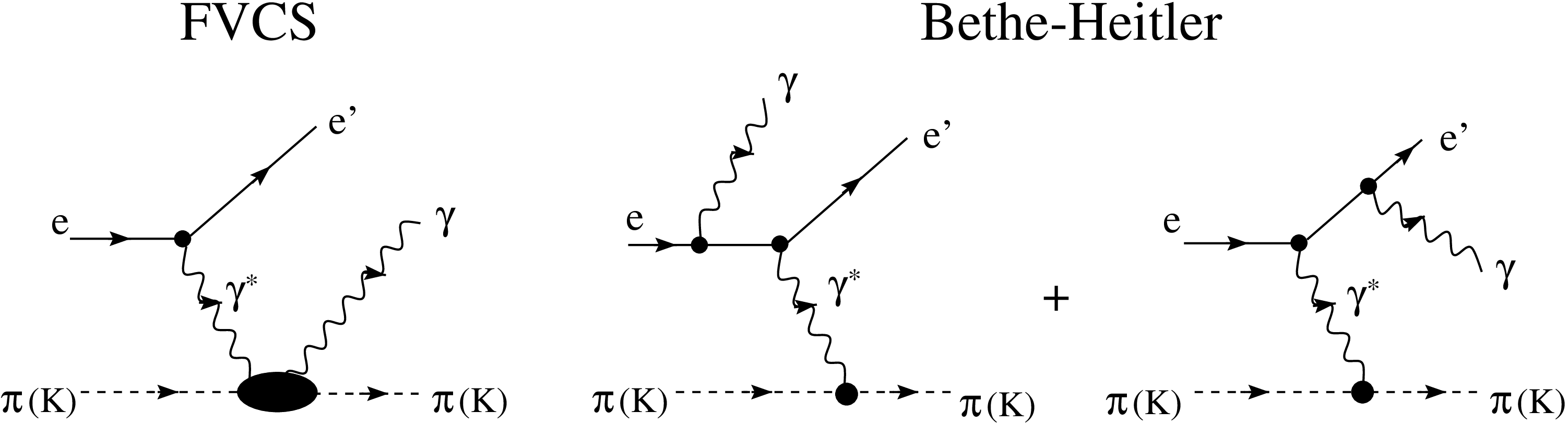, width=10cm, height=2.1cm}}
   \end{center}
\caption{\label{bhvcs} Bethe-Heitler and FVCS diagrams.}
\end{figure}

VCS off the pseudoscalar meson octet can be accessed experimentally via the 
electron-meson bremsstrahlung reaction. 
Such events are presently under analysis in the SELEX-E781 experiment
\cite{Selex}, where negative high energy pions of $\sim$ 600 GeV are
inelastically scattered off atomic electrons,
$\pi^- (p) e^- (k) \to  \pi^- (p') e^- (k') \gamma (q').$
The amplitude for this reaction is given by the sum of the Bethe-Heitler (BH)
and the full virtual Compton scattering (FVCS) contributions,
${\cal M} = {\cal M}_{\rm BH} + {\cal M}_{\rm FVCS}$ (see Fig. \ref{bhvcs}).

The BH diagrams correspond to the emission of the final photon
from the electron in the initial or final states, and involves only on-shell
information of the target, like the mass, the charge and the
 electromagnetic form factor. 
The explicit expression for the BH term reads
  \begin{eqnarray}
    {\cal M}_{\rm BH} &=&
      - \frac{{\rm i}e^3\epsilon^{\prime\ast}_{\mu}}{r^2}\bar{u}(k',s_{k'})
      \left(
      \frac{\gamma^{\mu}q' \cdot \gamma + 2k^{\prime\mu}}
      {2k'\cdot q'}\gamma^{\nu}
      + \gamma^{\nu}\frac{ q' \cdot \gamma \gamma^{\mu} - 2k^{\mu}}
      {2k \cdot q'} \right) u(k,s_{k}) \nn \\ &&
      \times \left(p+p'\right)_{\nu}F_{\pi}(r^2),
  \end{eqnarray}
where 
$r^\mu = p^\mu - p^\prime\,^\mu$ is the four-momentum transfer to the target.

In the one-photon exchange approximation, the FVCS amplitude is
proportional to the VCS amplitude, and reads
  \begin{eqnarray}
    {\cal M}^{\rm FVCS}&=&-\frac{{\rm i}e^3}{q^2}\bar{u}(k',s_{k'})\gamma_{\mu}
    u(k,s_k)\epsilon'_{\nu}\,^{*} {\cal M}_{\rm VCS}^{\mu\nu}.
  \end{eqnarray}
The differential cross section is then obtained from the coherent sum of the BH
and FVCS contributions. In the limit $\omega'\rightarrow 0$,
the interference of the BH and Born terms with the leading
contribution of the residual VCS amplitude contains information on the
generalized polarizabilities.

\begin{figure}[ht]
  \begin{center}
     \mbox{\psfig{file=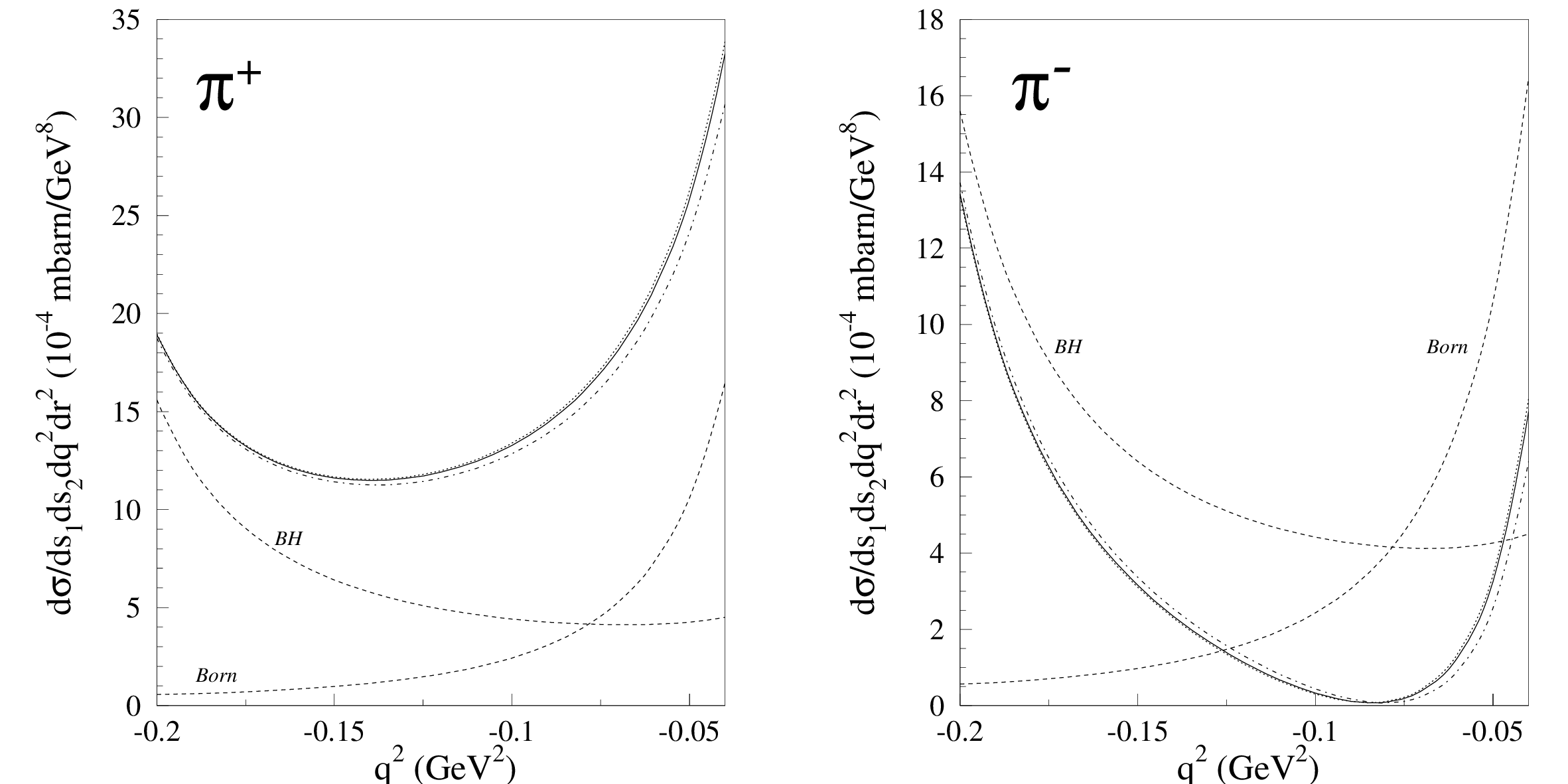,height=5.47cm}} 
  \end{center}
\caption{\label{pioncs} Differential cross sections for the charged pions.
Full lines: total results;
dashed lines: BH and Born contributions;
dotted lines: background contribution corresponding to the coherent sum of
the BH and Born terms;
dashed-dotted lines: total results with the residual VCS contribution
calculated without taking into account the $q^2$ dependence of the 
polarizability and using $\alpha_\pi=10\times 10^{-4}\, \mbox{fm}^3$.}
\end{figure}

In Fig. \ref{pioncs} we show the differential cross section for the charged
pions in a kinematical regime which can be accessed by the SELEX
experiment \cite{Selex}: $s=(p+k)^2=0.68\,\mbox{GeV}^2$,
$s_1=(k'+q')^2=0.154\,\mbox{GeV}^2$, $s_2=(p'+q')^2=0.2117\,\mbox{GeV}^2$,
$r^2=(p-p')^2=-0.0947\,\mbox{GeV}^2$.
The main contribution to the total cross section is given by the interference
between the Born and the BH terms.
This interference term is positive for $\pi^+$ and negative for
$\pi^-$ since the BH term is proportional to the pion form factor.
The effects of the polarizabilities correspond to the difference between
the total result (solid  lines) and the background contribution of
the BH and Born terms (dotted lines).
In addition, the effects due to the $q^2$ dependence
of the polarizability can be seen by comparing the full lines with the
dashed-dotted lines, where the contribution of the residual
VCS amplitude is calculated using the constant value 
of $\alpha_\pi= 10\times 10^{-4}\, \mbox{fm}^3$.

\section{Summary}
\label{section6}
We have calculated the invariant amplitude for VCS off the pseudoscalar meson
octet in ChPT at ${\cal O}(p^4)$.
This amplitude was split into a generalized Born and an non-Born term,
where each term was separately gauge invariant. 
In the limit $\omega' \rightarrow 0$, the non-Born contribution 
can be parametrized in terms of generalized polarizabilities. 
We used for the definition of these quantities a covariant approach 
interpreted in the pion Breit frame.
In the framework of ChPT at ${\cal O}(p^4)$, we found that the
momentum dependence of the generalized kaon and pion polarizabilities
is entirely given in terms of the mass and the decay constant of the meson.
Finally, we have investigated the possibility to extract information about the
generalized pion polarizabilities through the electron-pion
bremsstrahlung reaction.
This reaction receives a large background contribution from the
Bethe-Heitler and Born terms, and very high precision measurements are
necessary to disentangle the effects of the polarizabilities.

\begin{appendix}
\section{The loop function ${\cal G}(m^2,q^2,q \cdot q')$}
The function ${\cal G}(m^2,q^2,q \cdot q')$ is defined via
  \begin{equation}
     {\cal G}(m^2,q^2,q\cdot q') =
     1 + \frac{m^2}{q\cdot q'}\left[J^{(-1)}(a)-J^{(-1)}(b)\right]
     - \frac{q^2}{2q\cdot q'}\left[J^{(0)}(a)-J^{(0)}(b)\right],
  \end{equation}
where
  \begin{displaymath}
    J^{(n)}(x) := \int_0^1 {\rm d}y y^n\ln[1+x(y^2-y)-{\rm i}0^+]
  \end{displaymath}
and
  \begin{displaymath}
     a := \frac{q^2}{m},\quad b := \frac{q^2-2q\cdot q'}{m^2}.
  \end{displaymath}
The functions $J^{(0)}$ and $J^{(-1)}$ can be expressed as
  \begin{eqnarray*}
    J^{(0)}(x) &=& \left \{ \begin{array}{ll}
    -2 - \sigma\ln\left(\frac{\sigma-1}{\sigma+1}\right) & (x<0),\\
    -2 + 2\sqrt{\frac{4}{x}-1}\,\mbox{arccot}
    \left(\sqrt{\frac{4}{x}-1}\right) & (0 \le x < 4),\\
    -2 - \sigma\ln\left(\frac{1-\sigma}{1+\sigma}\right)-{\rm i}\pi\sigma
    & (4<x),
    \end{array} \right.\\
    J^{(-1)}(x) &=& \left \{ \begin{array}{ll}
    \frac{1}{2}\ln^2\left(\frac{\sigma-1}{\sigma+1}\right) & (x< 0),\\
    -\frac{1}{2}\arccos^2\left(1-\frac{x}{2}\right) & (0\le x <4),\\
    \frac{1}{2}\ln^2\left(\frac{1-\sigma}{1+\sigma}\right) - \frac{\pi^2}{2}
    + {\rm i}\pi\ln\left(\frac{1-\sigma}{1+\sigma}\right) & (4< x),
    \end{array} \right.
  \end{eqnarray*}
with
  \begin{displaymath}
    \sigma(x)=\sqrt{1-\frac{4}{x}},\quad x\notin [0,4].
  \end{displaymath}
The function $J^{(0)'}(x)$ is given by
  \begin{equation}
    \label{J0p}
    J^{(0)'}(x) = \frac{1}{x} \left[ 1 - 
    \frac{2}{x\sigma}\ln\left( \frac{\sigma-1}{\sigma+1}\right) \right] = 
    \frac{1}{x}\left[ 1 + 2J^{(-1)'}(x) \right], \qquad x<0.
  \end{equation}

\end{appendix}

\end{document}